\documentclass[final,1p,times]{elsarticle}
\usepackage{color}
\usepackage{bm}

\usepackage{graphicx}
\usepackage{amsmath}

\usepackage{amssymb}
\usepackage{gensymb}
\newcommand{\Alfven}{Alfv\'en }
\renewcommand{\div}{\bm{\nabla} \cdot}
\newcommand{\rot}{\bm{\nabla}\times}

\renewcommand{\Lambda}{\varLambda}
\renewcommand{\Gamma}{\varGamma}
\renewcommand{\Pi}{\varPi}

\newif\ifhighlight
\highlightfalse
\ifhighlight
\newcommand{\blue}{\color{blue}}
\newcommand{\red}{\color{red}}

\newcommand{\magenta}{\color{magenta}}
\newcommand{\black}{\color{black}}

\else
\newcommand{\blue}{}
\newcommand{\red}{}

\newcommand{\magenta}{}
\newcommand{\black}{}

\fi

\journal{Journal of Computational Physics}

\begin{document}

\begin{frontmatter}



%\title{Energetic-Particle-Magnetohydrodynamics Hybrid Simulation: A
%Generalized Quasi-Neutral Fluid-Particle Hybrid Plasma Model}

\title{A Generalized Quasi-Neutral Fluid-Particle Hybrid Plasma Model and Its
Application to Energetic-Particle-Magnetohydrodynamics Hybrid Simulation}


\author{Takanobu Amano\corref{corresponding}} \ead{amano@eps.s.u-tokyo.ac.jp}
\cortext[corresponding]{Corresponding author}

\address{Department of Earth and Planetary Science, University of Tokyo,
Tokyo, 113-0033, Japan}

\begin{abstract}
A generalized fluid-particle hybrid model for collisionless plasmas under the assumption of quasi-neutrality is presented. The system consists of fluid ions and electrons as well as arbitrary numbers of species whose dynamics is governed by the Vlasov equation. The proposed model is thus a generalized version of the well-known standard hybrid plasma simulation model, in which the ions are fully kinetic whereas the electrons are assumed to be a fluid. Since the proposed model employs the exact form of the generalized Ohm's law, the mass and energy densities, as well as the charge-to-mass ratio of the kinetic species, are taken to be arbitrary. In the absence of the kinetic species, it reduces to the quasi-neutral two-fluid model \cite{Amano2015}. In the opposite situation where the mass and energy densities of the kinetic species are much larger than the fluid ions, it is nothing more than the standard hybrid model with finite electron inertia effect. If the kinetic species is an energetic particle (EP) population having negligible mass density but substantial energy density and the scale size is much larger than the ion and electron inertial lengths, it describes the self-consistent coupling between the magnetohydrodynamics (MHD) and the EP dynamics. The energetic-particle-magnetohydrodynamics (EP-MHD) hybrid model is thus a special case of the more general model described in this paper. Advantages of this approach over the existing models are discussed. A three-dimensional simulation code solving the proposed set of equations is described. The code combines the Particle-in-Cell scheme for solving the kinetic species and a Riemann-solver-based code for the two-fluid equations. Several benchmark simulation results are shown to confirm that the code successfully captures the dynamics of the EP population interacting self-consistently with the MHD fluid.
\end{abstract}


\end{frontmatter}

%\linenumbers

\section{Introduction}
\label{intro}

Collisionless space, astrophysical, and laboratory plasmas are known to exhibit a rich variety of phenomena with a wide range of temporal and spatial scales. The standard magnetohydrodynamics (MHD) description has been quite successful in modeling macroscopic dynamics. However, it is well recognized that microphysics sometimes plays key roles in many practical applications. Numerical modeling of collisionless plasmas thus requires multiscale and/or multiphysics aspect.

For example, it has been known that a finite temperature anisotropy exists quite frequently in collisionless space plasmas. In fact, even if the system is initially isotropic, the anisotropy may easily be produced by adiabatic compression or expansion of a fluid element in an inhomogeneous magnetic field. Since the effect of temperature anisotropy is not negligible even at the MHD scale, it is one of the important physics that needs to be included for modeling large-scale phenomena. The Chew-Goldberger-Low (CGL) model \cite{Chew1956} has been the standard way to take into account the anisotropy in the MHD. Based on the CGL-MHD, a more sophisticated heat flux model for the collisionless plasma was invented \cite{Snyder1997}. Advanced shock-capturing numerical schemes for the CGL-MHD have also been developed recently \cite{Meng2012,Hirabayashi2016}.

Another important effect missing in the ideal MHD approximation is that of non-thermal energetic particles (EPs). Despite the fact that the presence of EPs is a quite common feature in many environments, the roles played by EPs cannot properly be taken into account in the ideal MHD model. The EP population usually has a negligible contribution to the density, whereas their energy density can be substantial. Therefore, they may have non-negligible contribution to the plasma pressure and affect the overall dynamics even at macroscopic scales. Indeed, the dynamical scale size of the EPs can be much larger than the kinetic scale of the thermal ions and electrons. One may assume that characteristic spatial scale sizes of the EPs are given by their Larmor radii. Since, by definition, the Larmor radii of the EPs are much larger than the thermal population, the effect of EPs may become significant even at large scale where the MHD approximation would be adequate in the absence of the EP population. This indicates that the self-consistent coupling between the background cold plasma and the EP population needs to be appropriately taken into account even at the MHD scale. We call such a model as the Energetic-Particle-Magnetohydrodynamics (EP-MHD) hybrid model in this paper.

There are quite a few examples in which the EP effect is believed to be important. It is well known that the cosmic rays (CRs) in the Galaxy have an average energy density near equipartition with the thermal gas as well as the magnetic field. (Note that the terms CR and EP are used interchangeably in this paper.) Therefore, the CRs play important roles in the dynamics of the interstellar medium. Also, the production rate of CRs at a strong shock is believed to be regulated by themselves through the feedback effect of the CR acceleration, and considerable effort has been made to understand the nonlinear shock acceleration efficiency \cite{Drury1981}. In space physics, EPs are commonly observed associated with, for instance, violent solar flares and/or coronal mass ejections. Estimated energies contained in the accelerated particles are often substantial fractions of the total energy released in these phenomena \cite{Miller1997,Mewaldt2005}. The solar wind disturbances interacting with the terrestrial magnetosphere sometimes cause a geomagnetic storm, which is caused by an enhanced ring current flowing westward near the equatorial plane at distances of several Earth radii. The major part of the ring current is carried by protons and oxygens with typical energies of $1\rm{-}100$ keV injected from the magnetotail \cite{Daglis1999}. In the same region, there exists a cold ($\lesssim 1$ eV) plasma population originating from the ionosphere as well. Typically the cold plasma dominates the plasma density, whereas the hot ring current population dominates the plasma pressure. This gives another example in which the EP effect becomes important for the MHD scale dynamics. Finally, the EP population is important also in magnetic confinement fusion plasmas. For instance, energetic alpha particles born in deuterium-tritium fusion reactions should be confined efficiently in the device for self-sustained operation of fusion reactors.

In the examples mentioned above, different physical models have been employed for numerical modeling. For instance, for the effect of CRs interacting with the background fluid, it is customary to adopt the so-called two-fluid model in which the CRs are approximated by yet another fluid \citep{Ryu2003,Kuwabara2004,Girichidis2016}. Although this approach takes into account the interaction between the two populations qualitatively, the kinetic effect associated with the CRs is completely missing. To circumvent the problem, a lot of models with various approximation levels have been suggested \citep[e.g.,][]{Zachary1986,Reville2012,Reville2013,Bai2015}. The most comprehensive model at present is perhaps the recent work by \cite{Bai2015} that takes into account the self-consistent coupling between fully kinetic CRs and the MHD fluid. We will show that the model is indeed very close to the one proposed in this paper. In magnetospheric physics, the ring current dynamics has been modeled essentially in a test-particle fashion \cite{Jordanova1997,Fok2001,Ebihara2003}. Such a model solves a particle transport equation in a prescribed electromagnetic field, but the feedback effect is not included. Recent effort has been devoted to combining such a conventional ring-current model with a global MHD simulation, trying to fill the gap between models based on different philosophies \cite{DeZeeuw2004,Pembroke2012,Glocer2013}. This is still an area of active research and the methodologies used in the ring-current modeling remain rather ad-hoc. We have previously proposed a numerical model in which the ring-current particle transport (approximated by the drift-kinetic equation) is solved in conjunction with the electromagnetic field (evolves according to Maxwell's equations) in a self-consistent fashion \cite{Amano2011}. On the other hand, this lacks the dynamics of the cold plasma population. Similarly, the transport of alpha particles in fusion plasmas has been modeled by the drift-kinetic equation. The current produced by the EPs is incorporated into the momentum conservation law in the MHD equations for the coupling between the two components \cite{Todo1995,Todo1998}. Instead of the drift-kinetic equation, the gyrokinetic equation may also be employed for solving the EPs coupled with the MHD equations \cite{Cheng1991,Belova1997}.

Despite the commonality for the presence of EPs, there is no standard method for numerical modeling of the self-consistent interaction between the EP and thermal populations. As we have seen above, different models have been employed in different communities. It is not easy to understand the range of applicability of the models used for different problems. It is the purpose of this paper to give a comprehensive view on how to incorporate the effect of EPs onto the MHD dynamics. In doing so, we find that it is rather straightforward to start from a more general model, in which the plasma consists of fluid ions and electrons, as well as arbitrary numbers of kinetic species having arbitrary mass densities, energy densities, and charge-to-mass ratios. Such a model indeed gives a generalization of the standard hybrid simulation model in which ions are solved in a fully kinetic way whereas electrons are approximated by a massless charge-neutralizing fluid. We show that some of the existing models may be obtained as special cases of the generalized hybrid model proposed in this paper. This thus clarifies the conditions under which the assumptions and approximations adopted in these models are reasonable.

We also present numerical methods for solving the proposed generalized set of equations. Since the model is essentially a combination of the fluid and kinetic parts, numerical methods for each component must appropriately be combined. The standard Particle-in-Cell (PIC) scheme is used for the kinetic part, whereas the fluid part is solved by a Riemann solver. We used the HLL-UCT (Harten-Lax-van Leer Upwind-Constrained-Transport) scheme \cite{Londrillo2004,DelZanna2007} that exactly satisfies the divergence-free condition for the magnetic field. These schemes have been implemented in a three-dimensional (3D) simulation code. The simulation result for test problems presented in Section \ref{sec:tests} confirms that the code captures the physics as designed.

This paper is organized as follows. Section \ref{sec:model} discusses the proposed physical model and how it is related to known models that have been used so far. The numerical methods are briefly described in Section \ref{sec:algorithm}. The results for several test problems are presented in Section \ref{sec:tests}. Finally, Section \ref{sec:summary} summarizes this paper.

\section{Model}
\label{sec:model}

\subsection{Basic equations}
\label{sec:basic_equations}

We start with the following two-fluid equations for the background ions and electrons with \blue the species index \black $s$ ($i$ for ions and $e$ for electrons):
\begin{align}
 & \frac{\partial}{\partial t} \rho_s +
 \div \left( \rho_s \bm{v}_s \right) = 0,
 \label{eq:basic_mass} \\
 & \frac{\partial}{\partial t} \rho_s \bm{v}_s +
 \div
 \left( \rho_s \bm{v}_s \bm{v}_s + p_s \bm{I} \right) =
 \frac{q_s}{m_s} \rho_s
 \left( \bm{E} + \frac{\bm{v}_s}{c} \times \bm{B} \right),
 \label{eq:basic_momentum} \\
 & \frac{\partial}{\partial t} \varepsilon_s +
 \div \left\{ \left( \varepsilon_s + p_s \right)
 \bm{v}_s \right\} = \frac{q_s}{m_s} \rho_s \bm{v} \cdot \bm{E},
 \label{eq:basic_energy}
\end{align}
where $q_s$, $m_s$, $\rho_s$, $\bm{v}_s$, $p_s$, \blue $\varepsilon_s$ \black are the charge, mass, mass density, bulk velocity, and (scalar) pressure (with $\bm{I}$ being the unit tensor), and total fluid energy density for the particle species $s$, respectively. The right-hand side represents the Lorentz force associated with the electromagnetic field $\bm{E}$, $\bm{B}$. The speed of light is denoted by $c$. A polytropic equation of state with a specific heat ratio $\gamma$ will be used throughout in this paper for the fluid species, \blue which gives the following relationship:
\begin{align}
 \varepsilon_s = \frac{1}{2} \rho_s \bm{v}_s^2 + \frac{1}{\gamma-1} p_s.
\end{align}
\black

The governing equation for the kinetic species is the standard Vlasov equation:
\begin{align}
 \frac{\partial}{\partial t} f_s +
 \bm{v} \cdot \frac{\partial}{\partial \bm{x}} f_s +
 \frac{q_s}{m_s}
 \left( \bm{E} + \frac{\bm{v}}{c} \times \bm{B} \right)
 \cdot \frac{\partial}{\partial \bm{v}} f_s =
 0,
 \label{eq:vlasov}
\end{align}
where $f_s (\bm{x}, \bm{v}, t)$ is the phase space density for the kinetic species $s$. The number of species, charge-to-mass ratio, mass density, energy density, etc., for the kinetic species are taken to be arbitrary. Indeed, a kinetic species is not necessarily a minor component. For instance, the mass density may even dominate the background fluid. We thus have to derive a model that does not assume anything, e.g., on the mass density contribution from the kinetic species.

We emphasize the distinction between the background fluid and kinetic species is arbitrary. A particle population may be treated as a kinetic species if one wants to take into account the kinetic effect associated with it. Otherwise, its density, velocity, pressure contributions may be pushed to the background fluid component. \blue Note that we could also have arbitrary numbers of fluid species. In principle, there is no technical difficulty in dealing with more than three fluid species. However, since our primary motivation is to develop a model which reduces to the EP-MHD hybrid model in a limiting case, we restrict ourselves to the two-fluid model for the fluid part in this paper. As we shall see below, this makes it possible to write the fluid equations in the conservative form for the total mass, momentum, and energy densities Eqs.~(\ref{eq:mass_conservation})-(\ref{eq:energy_conservation}). In the presence of more than three species, one may still use the three conservation laws but the fluid equations Eqs.~(\ref{eq:basic_mass})-(\ref{eq:basic_energy}) need to be solved for additional species.
\black

The charge and current densities may be calculated by taking the sum over all species (both fluid and kinetic species):
\begin{align}
 \varrho &=
 \sum_{s=i, e} \frac{q_s}{m_s} \rho_s +
 \sum_{s=kinetic} q_s \int f_s(\bm{v}) d \bm{v}
 \label{eq:charge}
 \\
 \bm{J} &=
 \sum_{s=i, e} \frac{q_s}{m_s} \rho_s \bm{v}_s +
 \sum_{s=kinetic} q_s \int f_s(\bm{v}) \bm{v} d \bm{v},
 \label{eq:current}
\end{align}
where the summation on the first term on the right-hand side represents contribution from the ion and electron fluids, whereas the second comes from the kinetic species. For later convenience, we define the charge and current densities of the kinetic species $\varrho_k$, $\bm{J}_k$ as follows
\begin{align}
 \varrho_{k}
 &=
 \sum_{s=kinetic} q_s \int f_s (\bm{v}) d \bm{v}
 \label{eq:kinetic_charge}
 \\
 \bm{J}_k
 &=
 \sum_{s=kinetic} q_s \int f_s (\bm{v}) \bm{v} d \bm{v}.
 \label{eq:kinetic_current}
\end{align}
In this paper we consider only low frequency phenomena in which the quasi-neutral approximation is adequate. Therefore, we always assume $\varrho \approx 0$ in the following.

Maxwell's equations in the low-frequency approximation (ignoring the displacement current) are given as follows:
\begin{align}
 & \frac{1}{c} \frac{\partial}{\partial t} \bm{B} = - \rot \bm{E}
 \label{eq:faraday}
 \\
 & \rot \bm{B} = \frac{4 \pi}{c} \bm{J},
 \label{eq:ampere}
\end{align}
with the divergence-free constraint for the magnetic field
\begin{align}
 \div \bm{B} = 0.
 \label{eq:divfree}
\end{align}
The validity of the low-frequency approximation was discussed by \cite{Amano2015} in conjunction with the quasi-neutrality assumption.

It is important to mention that the set of equations in the absence of the kinetic species is identical to the quasi-neutral two-fluid (QNTF) model described in \cite{Amano2015}. We have already shown that the model correctly reduces to the MHD in the long wavelength limit. On the other hand, in the limit of negligible background ion fluid density, it may be recognized as the standard hybrid with finite electron inertia effect appropriately taken into account. We will revisit this point later in Section \ref{sec:model_discussion}.

\subsection{Conservative form}
\label{sec:conservation_form}

As we have done in \cite{Amano2015} for the QNTF equations, it is convenient to cast the fluid equations into the conservative form. Considering finite contribution to current and charge densities from the kinetic species, the conservation laws may be written in the following form:
\begin{align}
 & \frac{\partial}{\partial t}
 \left[ \sum_{s=i,e} \rho_{s} \right] +
 \nabla \cdot
 \left[ \sum_{s=i,e} \rho_{s} \bm{v}_s \right]
 = 0. &
 \label{eq:mass_conservation}
 \\
 & \frac{\partial}{\partial t}
 \left[ \sum_{s=i,e} \rho_{s} \bm{v}_s \right] +
 \nabla \cdot
 \left[
 \sum_{s=i,e}
 \left( \rho_{s} \bm{v}_s \bm{v}_s + p_s \bm{I} \right) +
 \frac{\bm{B}^2}{8 \pi} \bm{I} -
 \frac{\bm{B} \bm{B}}{4 \pi}
 \right]
 =
 - \left(
 \varrho_k \bm{E} + \frac{\bm{J}_k}{c} \times \bm{B}
 \right), &
 \label{eq:momentum_conservation}
 \\
 & \frac{\partial}{\partial t}
 \left[ \sum_{s=i,e} \varepsilon_s +
 \frac{\bm{B}^2}{8\pi}
 \right] +
 \nabla \cdot
 \left[ \sum_{s=i,e}
 \left( \varepsilon_s + p_s \right) \bm{v}_s +
 c \frac{\bm{E} \times \bm{B}}{4\pi}
 \right]
 = - \bm{J}_k \cdot \bm{E}, &
 \label{eq:energy_conservation}
\end{align}
where we have made use of Maxwell's equations. The left-hand sides of the above equations are in the conservative form and may be understood as the mass, momentum, and energy conservation laws, respectively. The source terms on the right-hand sides represent the effect of the kinetic species. Namely, the momentum and energy gained (lost) by the kinetic species are compensated by the corresponding loss (gain) by the background fluid.

\subsection{Ohm's law}
\label{sec:ohms_law}

It is easy to show that the generalized Ohm's law may be written quite generally in the following form (see \ref{appendix:ohmslaw}):
\begin{align}
 \left(\Lambda + c^2 \nabla \times \nabla \times \right) \bm{E} = -
 \frac{\bm{\Gamma}}{c} \times \bm{B} + \div \bm{\Pi} + \eta
 \Lambda \bm{J}.
 \label{eq:ohm}
\end{align}
Here we have introduced a phenomenological resistivity $\eta$ for the sake of numerical convenience. The quantities $\Lambda$, $\bm{\Gamma}$, $\bm{\Pi}$ appearing in the above equation are defined with the moment quantities of both the fluid and kinetic species:
\begin{align}
 \Lambda &=
 \sum_{s=i, e} \frac{4 \pi \rho_s q_s^2}{m_s^2} +
 \sum_{s=kinetic} \frac{4 \pi q_s^2}{m_s} \int f_s d \bm{v}
 \label{eq:lambda} \\
 \bm{\Gamma} &=
 \sum_{s=i, e} \frac{4 \pi \rho_s q_s^2}{m_s^2} \bm{v}_s +
 \sum_{s=kinetic} \frac{4 \pi q_s^2}{m_s} \int \bm{v} f_s d \bm{v}
 \label{eq:gamma} \\
 \bm{\Pi} &=
 \sum_{s=i, e} \frac{4 \pi q_s}{m_s}
 \left( \rho_s \bm{v}_s \bm{v}_s + p_s \bm{I} \right) +
 \sum_{s=kinetic} 4 \pi q_s \int \bm{v} \bm{v} f_s d \bm{v}.
 \label{eq:pi}
\end{align}
This indeed gives a natural extension from the QNTF equations to the present system. Note that not only the zeroth and first order moments, the second order moment is also needed to calculate the electric field from the generalized Ohm's law. As we discuss (Section \ref{sec:model_discussion}) and actually demonstrate (Section \ref{sec:whistler_instability}) later, the second order moment indeed plays the central role if the EPs are electrons.

Again for convenience, we introduce the following symbols for the moment quantities associated with the kinetic species:
\begin{align}
 \Lambda_k &=
 \sum_{s=kinetic} \frac{4 \pi q_s^2}{m_s} \int f_s d \bm{v}
 \label{eq:lambda_k} \\
 \bm{\Gamma}_k &=
 \sum_{s=kinetic} \frac{4 \pi q_s^2}{m_s} \int \bm{v} f_s d \bm{v}
 \label{eq:gamma_k} \\
 \bm{\Pi}_k &=
 \sum_{s=kinetic} 4 \pi q_s \int \bm{v} \bm{v} f_s d \bm{v}.
 \label{eq:pi_k}
\end{align}
\blue
\subsection{Electron quantities}
\label{sec:electron_pressure}
\black

There have been no assumptions or approximations involved to derive the conservation laws and the generalized Ohm's law from the basic equations given in Section \ref{sec:basic_equations}. However, these equations complemented by Maxwell's equations are not sufficient to close the system because the electron pressure cannot be determined. \blue The electron density and velocity may be determined as (see \ref{appendix:standard_hybrid} for detail)
\begin{align}
 \rho_e &=
 \frac
 {\varrho_k + \frac{q_i}{m_i} D}
 {\frac{q_i}{m_i} - \frac{q_e}{m_e}}
 \\
 \bm{v}_e &=
 \frac
 {\bm{J}_k + \frac{q_i}{m_i} \bm{M} - \frac{c}{4 \pi} \nabla \times \bm{B}}
 {\rho_e (\frac{q_i}{m_i} - \frac{q_e}{m_e})},
\end{align}
where $D = \rho_i + \rho_e$ and $\bm{M} = \rho_i \bm{v}_i + \rho_e \bm{v}_e$ are the total mass and momentum densities obtained by updating the conservation laws Eqs.~(\ref{eq:mass_conservation}) and (\ref{eq:momentum_conservation}). \black On the other hand, it is only the total fluid (ion + electron) pressure that can be determined from the conservation laws, whereas the individual fluid pressures contribute differently in the generalized Ohm's law. Therefore, we have to introduce some assumption on the partition of the gas pressure between ion and electron fluids.

In conventional hybrid codes, the electron pressure is often determined by the polytropic equation of state $p_e \propto \rho_e^{\gamma}$, because they tend to behave adiabatically to fluctuations on the ion scale. Alternatively, one may solve a time evolution equation for the electron pressure (or entropy):
\begin{align}
 \frac{\partial}{\partial t} p_e + \nabla \cdot (p_e \bm{v}_e) =
 - (\gamma - 1) (\nabla \cdot \bm{v}) p_e
 + (\gamma - 1) \eta \bm{J}^2,
\end{align}
where the last term comes from irreversible Joule heating. In the previous work for the QNTF equations \cite{Amano2015}, we used the assumption that the electron and ion temperature ratio remains constant $\tau = T_i/T_e = {\rm const}.$ Although none of those approaches are correct in a strict sense, fortunately the evolution of the system is often insensitive to the choice of how the electron pressure is determined. For the test problems presented in this paper, we used the local polytropic equation of state $p_e \propto \rho_e^{\gamma}$.

\subsection{Model characteristics}
\label{sec:model_discussion}

As we have already mentioned, the present model reduces to the QNTF equations in the absence of the kinetic species, for which $\varrho_k$, $\bm{J}_k$, as well as the moment quantities $\Lambda_k$, $\bm{\Gamma}_k$, $\bm{\Pi}_k$ for the generalized Ohm's law are all negligible. This assures that the model correctly takes into account both the electron and ion (fluid) inertia effects but reduces to the MHD equations for scale sizes much larger than the ion inertial length.

With finite contribution from the kinetic species, the model in the long wavelength limit describes essentially the MHD equations self-consistently coupled with the dynamics of the kinetic species. This may be confirmed by the fact that the conservation laws in this regime are identical to those of the EP-MHD hybrid model proposed by \cite{Zachary1986}. They assumed the Ohm's law for the ideal MHD without much of discussion. This is indeed reasonable in realistic situations where the CRs are protons with a negligible density. However, this is not always the case, in particular with practical simulation parameters.

Remember that the major contributions to the Ohm's law come from light species as the moment quantities are inversely proportional to the mass. (This also applies to the second order moment as it is proportional to the pressure tensor multiplied by the charge-to-mass ratio.) Therefore, it is essentially the electrons that determine the Ohm's law, consistent with the familiar form of the Ohm's law used in the standard hybrid. There appears the Hall effect when the dynamics of the ion and electron fluids are decoupled because of different flow velocities between the two fluids. Similarly, \cite{Bai2015} found that there must be a correction term (which they called the CR-Hall term) to the ideal MHD Ohm's law when the decoupling between the thermal plasma and CRs is significant. This term tends to be important for a relatively large CR density. Such a large-density CR population may be adopted in practice to reduce computational costs of simulations. These effects are always included in the exact form of the Ohm's law. Admittedly, the remaining terms are small and usually unimportant corrections that can be ignored in most circumstances where the density of the kinetic species is negligible.

However, there are some advantages in using the exact form of the generalized Ohm's law. First, the density of the kinetic species can be arbitrary. If the density of the kinetic species is sufficiently small, it becomes the EP-MHD hybrid model. On the other hand, with a non-negligible density contribution from the kinetic species, its associated inertia effect is automatically included. In the extreme case where the density is dominated by the kinetic species, it becomes essentially the thermal population and the model is nothing more than the standard hybrid model (see, \ref{appendix:standard_hybrid}, how to recover the standard hybrid practically in a numerical code). Note that the finite electron inertia effect is included in the generalized Ohm's law in this case as well \cite{Amano2014}. The mass and momentum conservation laws become redundant equations that need not be solved if the fluid ion density is exactly zero. Nevertheless, we always solve these equations in our simulation code to handle general situations. Second, it enables us to deal with a kinetic population of an arbitrary charge-to-mass ratio. More specifically, if energetic electrons with a non-negligible energy density are treated as the kinetic species, their contributions to the Ohm's law must be included. In particular, we will show in Section \ref{sec:whistler_instability} that inclusion of the second order moment contribution is crucial for this case. Note that \cite{Bai2015} erroneously suggested that their formulation is applicable to energetic electrons as well. However, this is incorrect as is clear from our discussion. In general, one has to be careful to choose simulation parameters so as not to violate the assumptions made in constructing the numerical model. The generalized hybrid model resolves the issue as it always uses the exact form.

In summary, a numerical simulation code to solve the proposed model has quite general applications. It unifies the QNTF (including MHD), the classical hybrid with finite electron inertia, and the EP-MHD hybrid models. One subtlety is the distinction between the fluid and kinetic treatment. One has to separate the whole distribution into the thermal (fluid) and kinetic populations. Once this separation has been made, no kinetic effect can be included for the thermal population. Furthermore, energization of particles in the thermal population that may possibly inject them into ``the kinetic population'' is not taken into account unless some ad-hoc injection recipe is assumed. Nonetheless, the present model may become a useful tool to investigate the macroscopic dynamics of the system in which both the MHD fluid and pre-existing EPs play the roles.

\section{Numerical Algorithm}
\label{sec:algorithm}

As we have seen in the previous section, the present model may be considered as an extension of the QNTF model. The effect of kinetic species comes into the conservation laws for the QNTF as the external source terms on the right-hand sides. Modification needed for the generalized Ohm's law (to obtain the electric field) is just to add the contributions ($\Lambda_k$, $\bm{\Gamma}_k$, $\bm{\Pi}_k$) to the moment quantities, which can directly be computed by taking velocity moments of the distribution function. The magnetic field is advanced using Faraday's law. The electromagnetic field so obtained can also be used to solve the Vlasov equation (or the equation of motion) for the kinetic species. Therefore, our new code is built on top of the 3D QNTF code recently developed by \cite{Amano2015}.

In the following, we first summarize numerical methods for the fluid part used in the original QNTF code. A numerical algorithm for solving the kinetic species, and how the fluid and kinetic parts are combined together for temporal integration are then described.

\subsection{Fluid part}
\label{sec:fluid_part}

The QNTF code by \cite{Amano2015} solves the conservation laws coupled with Faraday's law for the magnetic field and the generalized Ohm's law for the electric field. It uses the HLL-UCT scheme \cite{Londrillo2004,DelZanna2007} which ingeniously combines the one-dimensional HLL Riemann solver and the UCT scheme for the magnetic induction equation in multidimensions. This satisfies the divergence-free condition up to machine precision while maintaining the shock-capturing capability of the Riemann solver. We have actually shown that the code is able to capture sharp MHD discontinuities if the resolution is insufficient to resolve the dispersive effect arising from the finite inertia of both ions and electrons. On the other hand, the dispersive effect is successfully reproduced with sufficiently fine resolution. Although our primary motivation is to develop an EP-MHD hybrid simulation code, the present model is indeed capable of dealing with Hall-MHD and even beyond.

Note that although the HLL scheme does not rely on the detailed spectral information of the system, it does require an estimate of the maximum wave propagation speeds (both in the positive and negative directions). We have used the linear dispersion relation for a homogeneous system to estimate the maximum wave speed in \cite{Amano2015}. In general, the eigenvalues of the system may differ in the presence of the kinetic species. It is actually neither easy nor practical to estimate and use the eigenvalues for the present system. We have tested several different choices on the estimate of the maximum wave speeds. One that works reasonably well and was used in the numerical examples in this paper is the simplest possible choice. That is, we have adopted the maximum phase speed $\alpha^{\pm}$ (with the sign indicating the positive/negative propagation direction) defined as
\begin{align}
 \alpha^{\pm} = {\rm max} \{ V_b \pm v_A, 0 \}
\end{align}
for the HLL Riemann solver, where $V_b$, $v_A$ are respectively the bulk velocity (defined as the center-of-mass velocity) and the \Alfven speed, both calculated using the local variables. Note that this is calculated for each direction with the corresponding component $V_b$, while $v_A$ is fixed. This assumes isotropy of the maximum wave propagation speed (i.e., independent of the magnetic field direction) which corresponds to the fast magnetosonic mode wave in the low-$\beta$ limit. It is thus smaller than the maximum wave propagation speed for a finite $\beta$ plasma or a plasma with the dispersive effect.  This rather crude estimate for the wave propagation speed is nonetheless adopted for the following reasons.

First, recall that the maximum wave speed in the HLL-type scheme controls the amount of dissipation in the scheme: The larger the wave speed, the stronger the numerical dissipation. Now, as we employ the standard PIC method for solving the kinetic species, the simulation always involves artificial noise due to insufficient numbers of particles. The thermal noise is generated by spontaneous emission due to charge and current density fluctuations associated with individual particle motions. Such thermal fluctuations may be re-absorbed by the particles, and eventually, the emission and absorption will balance with each other, resulting in a statistical equilibrium state. However, numerical dissipation introduced for the purpose of numerical stability breaks this balance because fluctuations generated by spontaneous emission are damped before being absorbed by the particles. The end result is numerical cooling of the particle distribution. On the one hand, one has to reduce the amount of numerical dissipation to minimize the artificial cooling, on the other hand, finite dissipation is necessary for numerical stability. Our choice of the phase speed is thus essentially the lower limit of the phase speed. Although it may not necessarily be strong enough to suppress spurious numerical oscillations at discontinuities, this will not be a significant issue as finite thermal noise is anyway inevitable in the PIC scheme. In our numerical experiments, we found that this choice of the wave propagation speed is sufficient for numerical stability, but at the same time, the artificial cooling is not so significant for practical use. \blue We should also note that the best choice for the phase speed, in general, will depend on numerical parameters such as spatial resolutions.  The strategy adopted here may thus be recognized as a reasonable compromise.  It is not necessarily the best solution and further investigation is clearly needed concerning this issue.
\black

\subsection{Kinetic part}
\label{sec:kinetic_part}

We use the standard PIC scheme for solving the Vlasov equation for the kinetic species. Namely, a finite number of computational particles are distributed in phase space which represent the particle distribution. They move in phase space continuously according to the equation of motion, whereas the field quantities are defined at the mesh points. Therefore, a proper data transfer between the particles and the field must be implemented. For simplicity, we always define the field quantities (electromagnetic field and moment quantities) to interact with the particles at the cell center. Note that the primary magnetic field is defined on the face center for the HLL-UCT scheme, but is also interpolated to the cell center (see \cite{Amano2015} for detail). The other quantities are defined originally at the cell center.

The velocity moments of the kinetic species are collected by taking weighted sum over the particles at the cell centers. A shape function $S(\bm{x})$ is used for weighting the moments. The zeroth, first, and second order velocity moments for the particle species $s$ denoted as $I_{s}^0$, $\bm{I}_s^1$, $\bm{I}_s^2$ at the position $\bm{r}$ are given by
\begin{align}
 I_s^{0} (\bm{r}) &=
 \sum_{j} m_s \, S(\bm{r} - \bm{x}_{s,j}),
 \label{eq:mom0}
 \\
 \bm{I}_s^{1} (\bm{r}) &=
 \sum_{j} m_s \bm{v}_{s,j} \, S(\bm{r} - \bm{x}_{s,j})
 \label{eq:mom1}
 \\
 \bm{I}_s^{2} (\bm{r}) &=
 \sum_{j} m_s \bm{v}_{s,j} \bm{v}_{s,j} \, S(\bm{r} - \bm{x}_{s,j})
 \label{eq:mom2}
\end{align}
respectively. Here ($\bm{v}_{s,j}$, $\bm{x}_{s,j}$) represents the phase-space coordinate of the $j$-th particle of the species $s$, and the summation is taken over all particles. Considering the symmetry of the second order moment $\bm{I}_s^2$, ten moments must be calculated in total for each species. Using these quantities, the charge and current densities $\varrho_k$, $\bm{J}_k$ as well as the moment quantities $\Lambda_k$, $\bm{\Gamma}_k$, $\bm{\Pi}_k$ are obtained as follows:
\begin{align}
 \varrho_k &=
 \sum_{s=kinetic} \frac{q_s}{m_s} I_s^0,
 \\
 \bm{J}_k &=
 \sum_{s=kinetic} \frac{q_s}{m_s} \bm{I}_s^1,
 \\
 \Lambda_k &=
 \sum_{s=kinetic} \frac{4 \pi q_s^2}{m_s^2} I_s^0,
 \\
 \bm{\Gamma}_k &=
 \sum_{s=kinetic} \frac{4 \pi q_s^2}{m_s^2} \bm{I}_s^1,
 \\
 \bm{\Pi}_k &=
 \sum_{s=kinetic} \frac{4 \pi q_s}{m_s} \bm{I}_s^2.
\end{align}
Typically, a three-point binomial filter is applied to all the moment quantities to eliminate artificial noise before calculating these quantities. \blue In other words, the filtering operation for a physical quantity $\psi_i$ (where $i$ being the grid index in one dimension) defined by
\begin{align}
 \tilde{\psi}_{i} =
 \frac{\psi_{i-1} + 2 \psi_{i} + \psi_{i+1}}{4}
\end{align}
is applied. For multidimensional filtering, the tensor product form of the above operation may be used as a straightforward extension.
\black

The electromagnetic field at the position $\bm{x}_{s,j}$ for updating the $j$-th particle velocity is calculated as
\begin{align}
 \bm{E} (\bm{x}_{s,j}) &= \int
 S(\bm{r} - \bm{x}_{s,j}) \, \bm{E} (\bm{r}) \, d \bm{r},
 \\
 \bm{B} (\bm{x}_{s,j}) &= \int
 S(\bm{r} - \bm{x}_{s,j}) \, \bm{B} (\bm{r}) \, d \bm{r}
\end{align}
using the same shape function. Since in reality the electromagnetic field is defined on the mesh points, the above integration should be replaced by the corresponding discrete sum over neighboring grid points.

In the test problems discussed in this study, we employed the second-order shape function (or quadratic spline) to reduce the numerical noise. \blue In the one dimensional case, the weighting factors for a particle position $x$ located within the $i$-th cell ($|x- r_i| \leq \Delta h/2$ where $\Delta h$ is the cell size) are given by
\begin{align}
 S (r_i - x) &=
 \frac{3}{4} - \left(\frac{x - r_i}{\Delta h}\right)^2 \\
 S (r_{i \pm 1} - x) &=
 \frac{1}{2}
 \left(\frac{1}{2} \pm \frac{x - r_{i}}{\Delta h} \right)^2
\end{align}
for grid indices $i$ and $i \pm 1$, respectively. Again, the tensor product was used for multidimensions.
\black

\subsection{Time integration}
\label{sec:time_integration}

We define both the particle position and velocity at integer time steps: the same as the fluid and field quantities. For instance, $\bm{x}_{s,j}^{n}$, $\bm{v}_{s,j}^{n}$ are the position and velocity at $n$-th time step $t = t^{n}$. This differs from typical full particle codes in which they are defined in a staggered manner for the leap-frog time integration scheme. Unfortunately, the equations to be solved cannot easily be represented in the leap-frog form. Therefore, our strategy to simplify the problem is to define all the primary variables at the same time step and use the mid-point rule for the time integration.

Temporal discretization with a time step of $\Delta t$ of the fluid conservation laws and the magnetic field induction equation may be symbolically written as follows:
\begin{align}
 \frac{\bm{U}^{n+1} - \bm{U}^{n}}{\Delta t} &=
 - \nabla \cdot \bm{F}
 \left(\bm{W}^{*}, \bm{E}^{*}, \bm{B}^{*}\right)
 + \bm{S}
 \left(
 \bm{E}^{*}, \bm{B}^{*}, \varrho_k^{*}, \bm{J}_k^{*}
 \right),
 \label{eq:update_fluid} \\
 \frac{\bm{B}^{n+1} - \bm{B}^{n}}{\Delta t} &=
 - \nabla \times \bm{E}^{*},
 \label{eq:update_bfield}
\end{align}
where the variables on the right-hand side with the superscript $^{*}$ are those defined at $t = (t^n + t^{n+1})/2 = t^n + \Delta t/2$. Here, $\bm{W} = \{\rho_e, \bm{v}_e, p_e, \rho_i, \bm{v}_i, p_i\}$ represents the fluid primitive variables, and $\bm{U} = \bm{U} (\bm{W}, \bm{B})$, $\bm{F} (\bm{W}, \bm{E}, \bm{B})$, $\bm{S} (\bm{E}, \bm{B}, \varrho_k, \bm{J}_k)$ are the conservative variables, corresponding fluxes, and source terms in the conservation laws, respectively. Note that the electric field is determined by solving the generalized Ohm's law \blue Eq.~(\ref{eq:ohm}) combined with Eqs.~(\ref{eq:lambda})-(\ref{eq:pi}). \black This procedure may also be written symbolically as
\begin{align}
 \bm{E}^{n} = \bm{G}
 \left(
 \bm{B}^{n}, \bm{W}^{n}, \bm{M} (\bm{Z}^n)
 \right),
 \label{eq:update_efield}
\end{align}
where $\bm{M} (\bm{Z}^n)$ indicates a set of moment quantities defined by Eqs.~(\ref{eq:mom0})-(\ref{eq:mom2}) as functions of the particle phase-space coordinate $\bm{Z}^n = \{\bm{v}_j^n, \bm{x}_j^n\}$.

Similarly, the particle positions and velocities are updated in the following way using the electromagnetic field defined at the intermediate step:
\begin{align}
 \begin{split}
 \frac{ \bm{x}_{j}^{*} - \bm{x}_{j}^{n} }{\Delta t / 2}
 &= \bm{v}_{j}^{n},
 \\
 \frac{ \bm{u}_{j}^{n+1} - \bm{u}_{j}^{n} }{\Delta t}
 &= \frac{q_j}{m_j}
 \left(
 \bm{E}^{*} (\bm{x}_{j}^{*}) +
 \frac{\bm{u}_j^{n+1} + \bm{u}_j^{n}}{2 \gamma_j^{n+1/2} c} \times
 \bm{B}^{*} (\bm{x}_{j}^{*})
 \right),
 \\
 \frac{ \bm{x}_{j}^{n+1} - \bm{x}_{j}^{*} }{\Delta t / 2}
 &= \bm{v}_{j}^{n+1}.
 \end{split}
 \label{eq:update_particle}
\end{align}
where the species index $s$ is omitted. In \magenta the above equations, $\bm{u}_j = \gamma_j \bm{v}_j$ with $\gamma_j = \sqrt{1 + \bm{u}_j^2/c^2}$ and $\bm{u}_j$ being the Lorentz factor and the four-velocity, respectively. We here take into account the relativistic effect, so that the code may also be used for modeling relativistic CRs. Note that, however, Maxwell's equations and the fluid equations do not consider the relativistic effect and both the flow and \Alfven speeds must be non-relativistic. \black In any case, the relativistic effect for the kinetic species is not important for the test problems presented in this paper.

The above procedure consists of three steps: (1) a half time step update of position, (2) a full time step update of velocity using the standard Buneman-Boris algorithm, (3) a half time step update of position. The electromagnetic fields $\bm{E}^{*} (\bm{x}_{j}^{*})$, $\bm{B}^{*} (\bm{x}_{j}^{*})$ are to be evaluated at the intermediate particle position $\bm{x}_{j}^{*}$. Given the intermediate electric and magnetic fields, it does not require a leap-frog type staggering of the variables in time. This is thus suitable to be combined with the fluid update. Overall, this gives a second-order accurate time integration scheme. A similar method has been used in Vlasov codes \cite{Cheng1976a,Mangeney2002} as well as in a standard hybrid code \cite{Kunz2014}.

We use the mid-point rule for evaluating the intermediate time step variables. Namely, $\bm{E}^{*} = (\bm{E}^{n} + \bm{E}^{n+1})/2$, $\bm{B}^{*} = (\bm{B}^{n} + \bm{B}^{n+1})/2$, and so forth. Although this makes the scheme formally implicit, a typical strategy is to use the predictor-corrector approach. In this study, we adopt the same procedure as suggested by \cite{Kunz2014}. The whole time integration procedure may be summarized as follows:
\begin{enumerate}
 \item Let the variables known at $n$-th step as initial guesses for the next
       time step: $\bm{B}_0^{n+1} = \bm{B}^{n}$, $\bm{W}_0^{n+1}=
       \bm{W}^{n}$, $\bm{Z}_{0}^{n+1} = \bm{Z}^{n}$, and thus,
       $\bm{E}_0^{n+1} = \bm{E}^n$.
 \item Update the magnetic field and fluid variables using
       Eqs.~(\ref{eq:update_fluid})-(\ref{eq:update_bfield}), which are then
       denoted as $\bm{B}_1^{n+1}$ and $\bm{W}_1^{n+1}$. The particles
       are not updated at this time and $\bm{Z}_1^{n+1} = \bm{Z}^{n}$
       remains unchanged. Update the electric field by $\bm{E}_1^{n+1} =
       \bm{G} (\bm{B}_1^{n+1}, \bm{W}_1^{n+1}, \bm{M}
       (\bm{Z}_1^{n+1}))$.
 \item Update again the magnetic field and fluid variables to obtain
       $\bm{B}_2^{n+1}$ and $\bm{W}_2^{n+1}$. Now update the particles
       using Eq.~(\ref{eq:update_particle}) to obtain
       $\bm{Z}_2^{n+1}$. The electric field is calculated by
       $\bm{E}_2^{n+1} = \bm{G} (\bm{B}_2^{n+1},
       \bm{W}_2^{n+1}, \bm{M} (\bm{Z}_2^{n+1}))$.
 \item Using the above as the final guesses, update the magnetic field, fluid
       variables and particles to obtain $\bm{B}^{n+1}$,
       $\bm{W}^{n+1}$, $\bm{Z}^{n+1}$. The electric field is then
       calculated by $\bm{E}^{n+1} = \bm{G}
       (\bm{B}^{n+1}, \bm{W}^{n+1}, \bm{M} (\bm{Z}^{n+1}))$.
\end{enumerate}

\blue
We note that the time step is restricted by the CFL (Courant-Friedrichs-Lewy) condition and the accuracy of resolving individual particle motions. Which of them actually imposes the major restriction depends mainly on the spatial resolution. For a typical standard hybrid simulation with a cell size on the order of the ion inertial length, both are more or less comparable. For a much finer resolution, the CFL condition with respect to the whistler wave determines the time step. As our model includes the finite electron inertia effect, there exists an upper bound ($\sim \sqrt{m_i/m_e} v_A / 2$) in the phase velocity. Interestingly, for a spatial resolution comparable to the electron inertial length, the CFL condition may be written as $\Omega_{ce} \Delta t \lesssim 1/2$, where $\Omega_{ce}$ is the electron cyclotron frequency. For a cell size much larger than the ion inertial length, the CFL condition is no longer important and the time step is sorely determined by the individual particle gyromotions. In principle, the time step for the particle and field can be chosen independently (as has been done so in some hybrid codes). However, we use a single time step in the current implementation for simplicity.
\black

\section{Tests}
\label{sec:tests}

In this section, we present simulation results for several test problems. Although the simulation code is fully 3D, reduced-dimensional simulations were performed with two grid points in the homogeneous direction. Note that the two grid points are the minimum because we used the second-order shape function. The boundary condition is always periodic in all three directions.

In the following, the index for species $i$ and $e$ are used to represent the fluid ion and electron components, whereas $k$ is used for the kinetic species. Unless otherwise stated, a ratio between \Alfven speed and the speed of light of $v_A/c = 10^{-4}$, a polytropic index of $\gamma = 5/3$, an ion-to-electron mass ratio of $m_i/m_e = 100$, and a resistivity of $\eta = 0$ were used. Other simulation parameters are denoted as follows: \blue $\Delta t$ for time step, $\Delta h$ for cell size, \black $N_{ppc}$ for a number of particles per cell, $N_{j}$ ($j = x, y, z$) a number of grid points in $j$ direction. Time and space are measured in units of the inverse ion cyclotron frequency $\Omega_{ci}^{-1}$, and the ion inertial length $\lambda_i = c/\omega_{pi}$, where $\omega_{pi}$ is the ion plasma frequency defined with the total ion number density. The velocity and magnetic field are respectively normalized to the \Alfven speed $v_A$ and the ambient magnetic field $B_0$.

\red
The list of simulations presented in this paper is summarized in Table \ref{table:simulation} with some numerical parameters. Detailed descriptions of the simulation setup and physical parameters can be found in each subsection below. \magenta In all the examples, the total energy conservation errors were no worse than 1\%. Typically errors in EP-MHD hybrid simulations were roughly ten times better than the standard hybrid simulation.
\black

\begin{table}[hbtp]
  \caption{Summary of simulations.}
  \label{table:simulation}
  \centering
  \begin{tabular}{lllllll}
    \hline
    Run ID &
    Model &
    Section &
    $\Omega_{ci} \Delta t$ &
    $\Delta h / c/\omega_{pi}$ &
    $(N_x, N_y, N_{ppc})$ \\
    
    \hline \hline
    Run-1A &
    standard &
    \ref{sec:linear_waves} &
    $5 \times 10^{-3}$&
    $0.1$ &
    $(1024, 2, 512)$ \\
    
    Run-1B &
    standard &
    \ref{sec:linear_waves} &
    $5 \times 10^{-3}$&
    $0.1$ &
    $(1024, 2, 512)$ \\

    Run-2A &
    standard &
    \ref{sec:electromagnetic_ion_beam_instability} &
    $1 \times 10^{-2}$ &
    $0.25$ &
    $(1024, 2, 512)$ \\
    
    Run-2B &
    EP-MHD &
    \ref{sec:electromagnetic_ion_beam_instability} &
    $1 \times 10^{-2}$ &
    $0.25$ &
    $(1024, 2, 512)$ \\
    
    Run-3A &
    EP-MHD &
    \ref{sec:whistler_instability} &
    $5 \times 10^{-5} $ &
    $0.025$ &
    $(1024, 2, 512)$ \\
    
    Run-3B &
    EP-MHD &
    \ref{sec:whistler_instability} &
    $5 \times 10^{-5} $ &
    $0.025$ &
    $(1024, 2, 512$) \\
    
    Run-4A &
    standard &
    \ref{sec:firehose_instability} &
    $2 \times 10^{-2}$ &
    $0.5$ &
    $(256, 256, 256)$ \\
    
    Run-4B &
    EP-MHD &
    \ref{sec:firehose_instability} &
    $2 \times 10^{-2}$ &
    $5.0$ &
    $(256, 256, 256)$ \\
    
    \hline
  \end{tabular}
\end{table}

\subsection{Linear waves}
\label{sec:linear_waves}

We first discuss the propagation characteristics of linear waves in a homogeneous plasma. To reproduce kinetic and dispersive effects appearing at around the ion inertial length or gyroradius, we set $q_i/m_i = 0$ for the fluid ions and \blue all the ions \black were represented by particles with $q_k/m_k = -(m_e/m_i) (q_e/m_e)$. The problem thus demonstrates the capability that the code can be used as a standard hybrid code.

\begin{figure}[h]
 \begin{center}
  \includegraphics[scale=0.5]{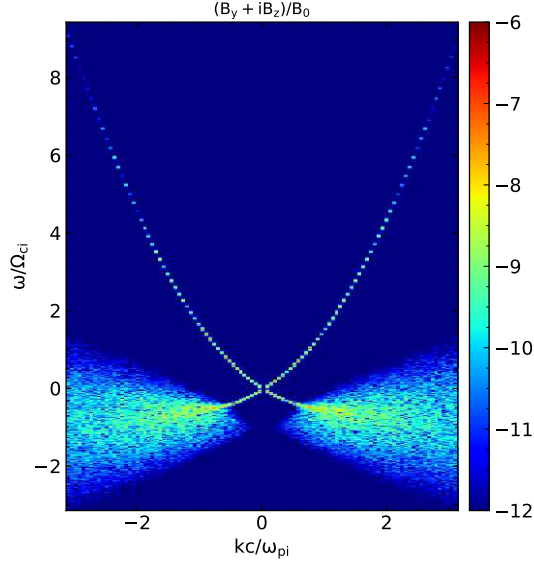}

  \caption{Power spectrum obtained by Run-1A for linear wave propagation
  parallel to the ambient magnetic field. The base-10 logarithm of the power
  calculated for the complex magnetic field $B_y + i B_z$ is shown in color.}

  \label{fig:linear_para}
 \end{center}
\end{figure}

The first simulation (Run-1A) was performed on a one-dimensional (1D) simulation box along the $x$ direction. The system was initially uniform, and the ambient magnetic field was directed along the $x$ direction. The other simulation parameters were chosen as follows: $\Omega_{ci} \Delta t = 5 \times 10^{-3}$, $\Delta h/c/\omega_{pi} = 0.1$, $N_x = 1024$, $N_{ppc} = 512$, $\beta_k = \beta_e = 0.1$. The plasma beta $\beta_{s}$ is defined as the ratio between the particle pressure and the magnetic pressure. The initial velocity distribution for the kinetic species is Maxwellian with a thermal velocity of $v_{thk}/v_A = \sqrt{\beta_k/2}$ (as defined by the standard deviation for the Gaussian distribution). The simulation started with initial noise inherent in the PIC scheme and was run up to $\Omega_{ci} t = 500$. The power spectrum was calculated by taking Fourier transform both in time and space of the series of snapshot data recorded every $\Omega_{ci} t = 0.25$ interval. The spectrum for the transverse magnetic field is shown in Fig.~\ref{fig:linear_para}. The result was obtained for the complex magnetic field $B_y + iB_z$. Thus, the positive (negative) wavenumbers indicate right-handed (left-handed) helicity, whereas positive (negative) frequencies indicate R-mode (L-mode) polarization. One can clearly see the whistler mode dispersion relation consistent with the cold plasma theory in the R-mode domain ($\omega > 0$). On the other hand, the L-mode \Alfven/ion-cyclotron waves are observed in the negative frequency range ($\omega < 0$). The enhanced power observed in the L-mode domain was due to fluctuations generated by doppler-shifted cyclotron motions of the thermal particles, and hence the L-mode waves were damped at high wavenumbers.

\begin{figure}[h]
 \begin{center}
  \includegraphics[scale=0.5]{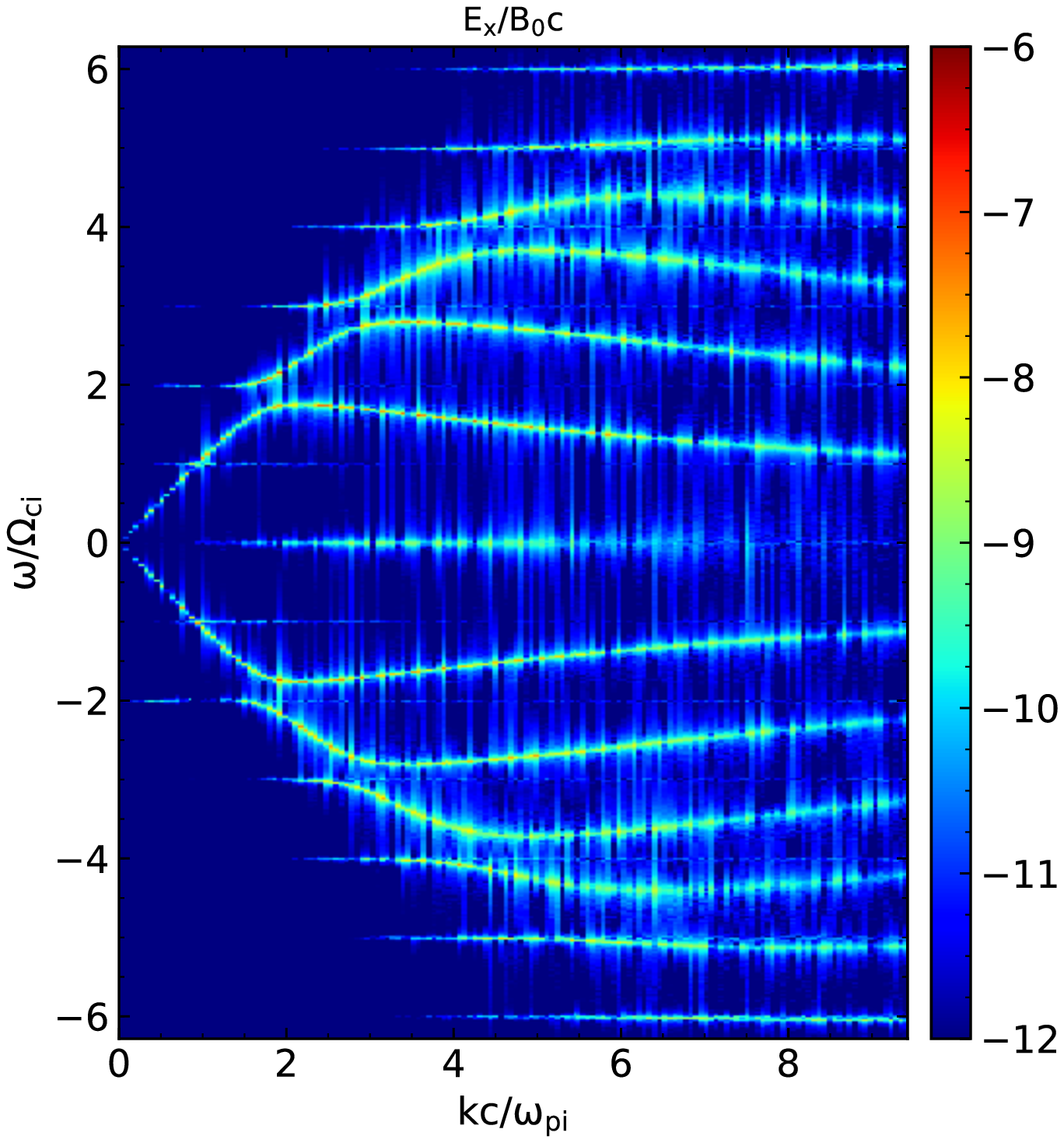}

  \caption{Power spectrum obtained by Run-1B for linear wave propagation
  perpendicular to the ambient magnetic field. The base-10 logarithm of the
  power calculated for the electrostatic field $E_x$ is shown in color.}

  \label{fig:linear_perp}
 \end{center}
\end{figure}

We then performed another simulation (Run-1B) with the ambient magnetic field pointing in the $z$ direction. The other simulation parameters were \blue the same as before \black. Fig.~\ref{fig:linear_perp} shows the power spectrum obtained for $E_x$. Low-frequency magnetosonic waves are observed at small wavenumbers, whereas harmonics structures or ion Bernstein waves are seen at large wavenumbers or frequencies. \red (The zero-frequency modes may be recognized as entropy modes.) \black The appearance of Bernstein waves clearly shows that the ion kinetic effect is correctly included. Similar results have been reported recently by \cite{Munoz2018}.

\subsection{Electromagnetic ion beam instability}
\label{sec:electromagnetic_ion_beam_instability}

In this subsection, we present 1D simulation results for an electromagnetic ion beam instability. This is one of the standard test problems for a hybrid simulation code and its linear and nonlinear evolutions are well understood. We consider an instability driven by a fast, dilute and cold beam propagating along the magnetic field, for which the most unstable mode appears in the MHD regime $k c/\omega_{pi} \ll 1$ via the cyclotron resonance of the beam component. The evolution of the system is therefore dominated by the kinetic effect of the beam ions whereas that of the background ions may not necessarily be important. Below we compare two simulation results: the one reproduced by the standard hybrid (Run-2A), the other adopts the EP-MHD hybrid model with the background ions being approximated by a fluid description (Run-2B).

We set up a 1D simulation box along the $x$ direction with a constant magnetic field applied also in the same direction. The ion velocity distribution function consists of two distinct populations; the core and beam components with a relative streaming speed along the ambient magnetic field. The beam density relative to the total (or electron) density was $n_b/n_0 = 0.02$, and the drift velocity for the beam and core components were $V_b/v_A = 9.8$ and $V_c/v_A =-0.2$, where the subscript $b$, $c$ indicate the beam and core, respectively. Notice that the drift velocities are chosen to satisfy the zero net current condition and the simulation frame corresponds to the rest frame of the electron fluid. \red Linear analysis for the instability indicates a maximum growth rate of $\gamma/\Omega_{ci} \simeq 0.21$ appears at a wavenumber of $k c/\omega_{pi} \simeq 0.12$. \black

For Run-2A, the both distributions are solved by particles, i.e., $q_i/m_i = 0$ and $q_k/m_k = -(m_e/m_i) (q_e/m_e)$ for $k = b, c$. Each velocity distribution is initialized by an isotropic Maxwellian distribution in its rest frame. The thermal velocities for the two components were assumed to be the same: $v_{thb}/v_A = v_{thc}/v_A = 1/\sqrt{2}$, giving $\beta_b = n_b/n_0 = 0.02$ and $\beta_c = 1 - n_b/n_0 = 0.98$, respectively. The electron plasma beta was $\beta_e = 1.0$. The other simulation parameters were as follows: $\Omega_{ci} \Delta t = 10^{-2}$, $\Delta h / c/\omega_{pi} = 0.25$, $N_x = 1024$, $N_{ppc} = 512$ (both for the beam and core components with different weights). In Run-2B, the core ion distribution was replaced by the ion fluid with the same plasma $\beta_i = 0.98$. The other parameters were the same as Run-2A.

\begin{figure}[h!]
 \begin{center}
  \includegraphics[scale=0.5]{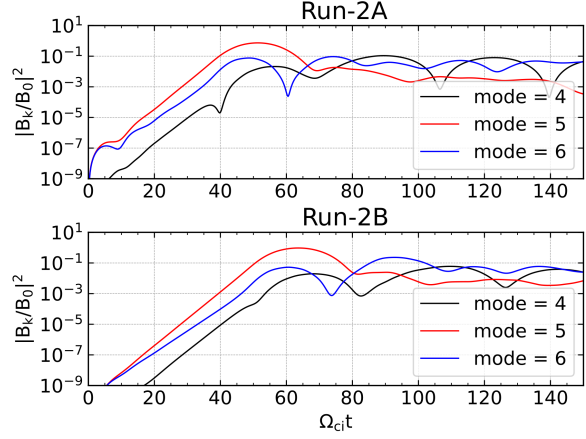}

  \caption{Time evolution of three Fourier mode amplitudes around the
  predicted maximum growth. The top and bottom panels show the results for
  Run-2A and Run-2B, respectively.}

  \label{fig:beam_growth}
 \end{center}
\end{figure}

\begin{figure}[h!]
 \begin{center}
  \includegraphics[scale=0.5]{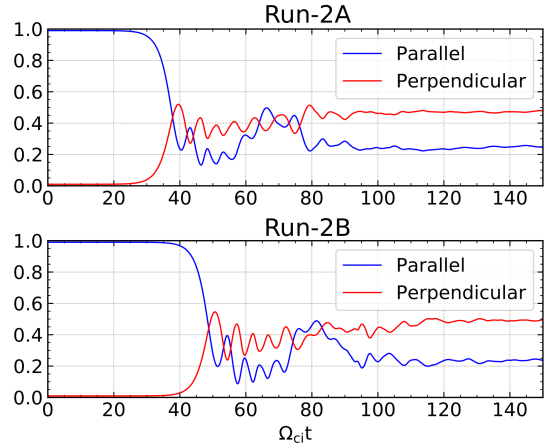}

  \caption{Time evolution of parallel and perpendicular energies for the beam
  component. The top and bottom panels show the results for Run-2A and Run-2B,
  respectively. The energy is normalized to the initial total beam energy.}

  \label{fig:beam_energy}
 \end{center}
\end{figure}

The growth of the instability in three Fourier modes (wavenumbers) at around the maximum growth is shown in Fig.~\ref{fig:beam_growth} for Run-2A (top) and Run-2B (bottom), respectively. The amplitudes were obtained as right-handed helicity modes by taking Fourier transform of the complex transverse magnetic field $B_y + i B_z$. Except for the initial noise level, the two simulation results (both the growth rates and saturation levels) agreed quite well with each other. The time history of parallel and perpendicular energies for the beam component are shown in Fig.~\ref{fig:beam_energy} for Run-2A (top) and Run-2B (bottom), in which the energy exchange between the two components due to the large amplitude wave and resulting pitch-angle scattering are clearly seen. One can see again \red quite good agreement \black between the two runs.

\begin{figure}[h!]
 \begin{center}
  \includegraphics[scale=0.8]{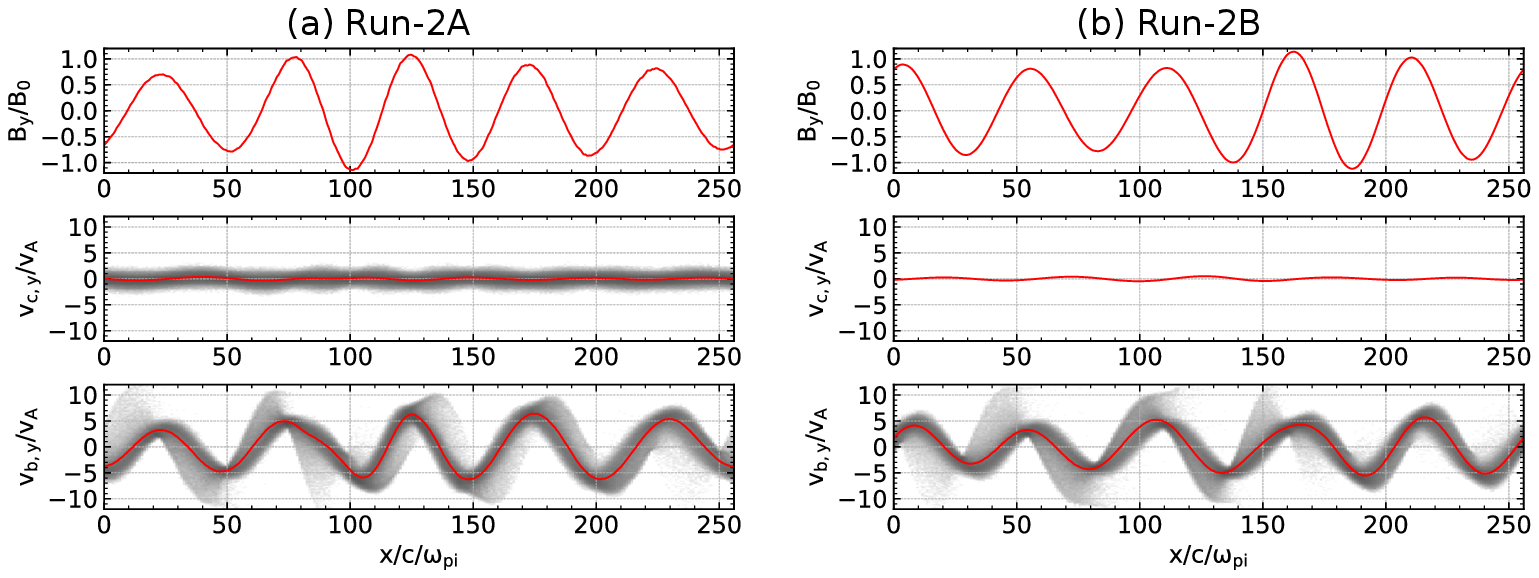}

  \caption{Comparison of snapshots between the standard (Run-2A) and EP-MHD (Run-2B) hybrid models. The left panel (a) displays a snapshot at $\Omega_{ci} t = 50$ for Run-2A, whereas a snapshot at $\Omega_{ci} t = 60$ is shown for Run-2B in the right panel (b). The magnetic field $B_y$ (top), and the moment velocity $v_y$ for the core (middle), and the beam (bottom) components are shown respectively with the solid lines. The phase space plots in $x$-$v_y$ space are also shown for those components solved by using particles.}
  \label{fig:beam_snapshot}
 \end{center}
\end{figure}

To compare kinetic wave-particle interaction behaviors, the ion phase space in $x$-$v_y$ space and moment velocities $v_y$ for the beam and core components together with the magnetic field profile $B_y$ are shown in Fig.~\ref{fig:beam_snapshot} (a) for Run-2A at around the saturation ($\Omega_{ci} t = 50$). A similar plot for Run-2B (with the core phase-space plot replaced by the fluid bulk velocity in the middle panel) is shown in Fig.~\ref{fig:beam_snapshot} (b) at approximately the same time around the saturation $\Omega_{ci} t = 60$. The resonant interaction of the beam component is clearly seen in both cases; the large amplitude oscillations of the velocity in phase with the magnetic field fluctuations. In contrast, the velocity fluctuation amplitudes for the core are much smaller.

All these results are consistent with the idea that the dynamics of the core component can reasonably be modeled with the fluid description. If one adopts the fluid description for the core component, the numerical noise associated with the use of macro-particles is greatly reduced, which thus improves the overall accuracy of the result.

\subsection{Whistler instability}
\label{sec:whistler_instability}

Now we demonstrate that the code correctly includes the kinetic effect of energetic electrons. We consider the whistler instability excited by a perpendicular temperature anisotropy $T_{\perp}/T_{\parallel} > 1$ of the energetic electron population. As the instability is driven unstable by the cyclotron resonance, it should be reproduced correctly only when the kinetic effect of the energetic electron component is appropriately taken into account.

To single out the essential physics, we used the EP-MHD hybrid model; the background ions and electrons were approximated by fluids, and only the energetic electrons were solved by particles with $q_k/m_k = q_e/m_e$. We considered a homogeneous plasma and the density of the energetic electrons $n_{k}$ was set to $n_{k}/n_0 = 0.01$ where $n_0$ is the total density. For charge neutrality, the background electron and ion densities were given by $n_e = n_0 - n_{k}$ and $n_i = n_0$. A temperature anisotropy of $T_{k,\perp}/T_{k,\parallel} = 9$, and a parallel beta of $\beta_{k,\parallel} = 0.1$ were used for the energetic electron component. The parallel and perpendicular thermal velocities were thus given by $v_{k,\parallel}/v_A \simeq 22.4$ and $v_{k,\perp}/v_A \simeq 67.1$, respectively. The plasma $\beta$ for the background ion and electron fluids were set as $\beta_i = 0.1$ and $\beta_e = \beta_i (1 - n_k/n_0) = 0.099$, so that the fluid pressures were comparable to the (parallel) pressure of the energetic population.

We used a 1D simulation box along the $x$ direction parallel to the ambient magnetic field. The simulation referred to as Run-3A used the following parameters: $\Omega_{ci} \Delta t = 5 \times 10^{-5}, \Delta h / c/\omega_{pi} = 2.5 \times 10^{-2}, N_x = 512, N_{ppc} = 1024$. Notice that the time step was chosen to be small enough to resolve the cyclotron motion of electrons (rather than protons). Similarly, the cell size was chosen to be much smaller than the ion inertial length because the instability appears at around a wavelength comparable to the electron inertial length $\lambda_e = \lambda_{i} / \sqrt{m_i/m_e}$.

\begin{figure}[h]
 \begin{center}
  \includegraphics[scale=0.5]{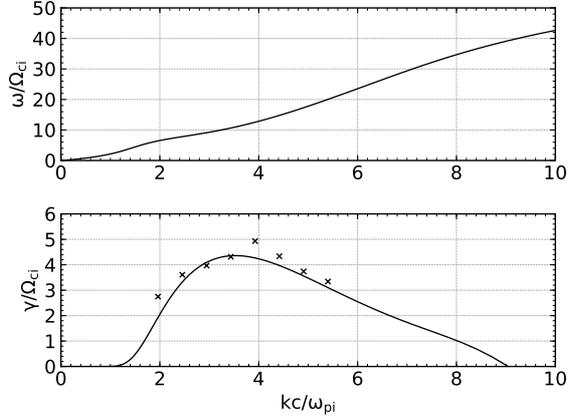}

  \caption{Result of linear dispersion analysis. The real frequency and the
  growth rate are shown in the top and bottom panels, respectively. In the
  bottom panel, the growth rates estimated by fitting the growth curve for
  each mode obtained by Run-3A during the linear growth phase are shown with
  crosses.}

  \label{fig:whistler_growth}
 \end{center}
\end{figure}

In Fig.~\ref{fig:whistler_growth}, the result of linear dispersion analysis is summarized. The growth rate becomes maximum \red ($\gamma/\Omega_{ci} \simeq 4.4$) \black at $k c/\omega_{pi} \simeq 3.4$ with a real frequency much larger than the ion cyclotron frequency $\omega/\Omega_{ci} \gg 1$. This confirms that the whistler mode branch is unstable. In the lower panel, the growth rates measured from the simulation results are also shown with crosses. This clearly shows that the simulation quite well reproduced the kinetic effect of the energetic electrons.

\begin{figure}[h]
 \begin{center}
  \includegraphics[scale=0.5]{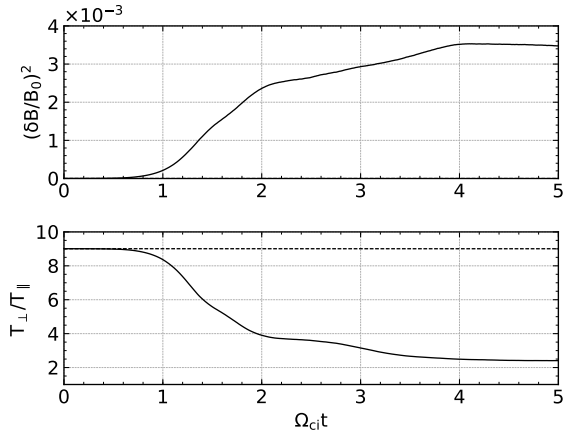}

  \caption{Time evolution of the magnetic field fluctuations (top) and the
  temperature anisotropy (bottom). The solid lines are results obtained by the
  fiducial code with the off-diagonal terms of $\bm{\Pi}_k$ included
  (Run-3A). On the other hand, the dashed lines are for the case without the
  off-diagonal terms (Run-3B). In the top panel, the dashed line is not
  visible because it is essentially due to thermal fluctuations and too small
  in this scale.}

  \label{fig:whistler_history}
 \end{center}
\end{figure}

\begin{figure}[h]
 \begin{center}
  \includegraphics[scale=0.5]{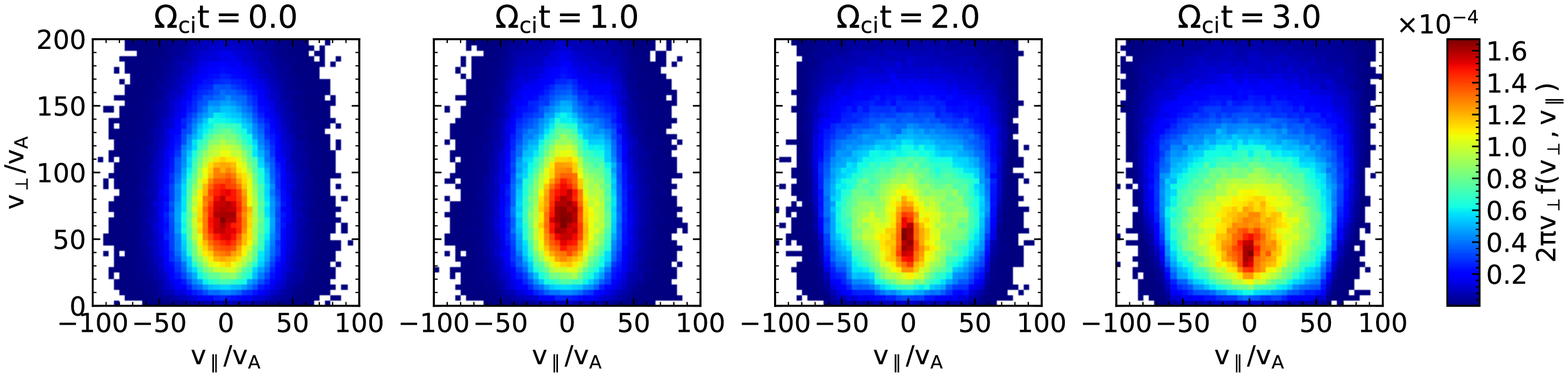}

  \caption{Time evolution of velocity distribution function obtained by
  Run-3A. Four snapshots of the velocity distribution function $2 \pi
  v_{\perp} f(v_{\perp}, v_{\parallel})$ integrated over the entire simulation
  box are shown in color. The distribution function is normalized such that
  the integral over velocity space becomes unity.}

  \label{fig:whistler_vdist}
 \end{center}
\end{figure}

The time history of the wave energy and the anisotropy are shown in Fig.~\ref{fig:whistler_history}. As the instability develops, the temperature anisotropy was clearly reduced as a result of pitch-angle scattering via the cyclotron resonance. Shown in Fig.~\ref{fig:whistler_vdist} is the evolution of the velocity distribution function $2 \pi v_{\perp} f(v_{\perp}, v_{\parallel})$ where the pitch-angle scattering of electrons by the whistler waves is evident. All these development of the instability is consistent with previous PIC simulation results obtained for the same instability but driven by a thermal electron anisotropy \citep[e.g.,][]{Devine1995,Gary1996}.

We should also mention that the growth of whistlers was not observed when we artificially dropped the off-diagonal terms of the $\bm{\Pi}_k$ tensor in the generalized Ohm's law (which is called Run-3B). Actually, the results without the off-diagonal terms are also plotted in Fig.~\ref{fig:whistler_history} with dashed lines (although not visible in the top panel). One can see that no relaxation of temperature anisotropy occurred and the electrons appeared to behave as if they had been test particles. This confirms that the use of the exact form of the Ohm's law is crucial, and any previous studies used approximate versions could not properly include the kinetic effect of energetic electrons.

\subsection{Firehose instability}
\label{sec:firehose_instability}

As the final test problem, we consider the firehose instability (FHI) driven by a parallel temperature anisotropy ($T_{\perp}/T_{\parallel} < 1$) \red of protons \black in a two-dimensional (2D) computational box. It has been well-known that FHI has actually two distinct modes. The so-called parallel FHI has a maximum growth rate at a wavenumber along the ambient magnetic field, whereas there appears another instability at oblique propagation which we call as the oblique FHI \cite{Hellinger2000}. The parallel FHI is on the magnetosonic/whistler wave branch and has a finite real frequency \cite[e.g.][]{Gary1998}, whereas the oblique FHI is purely growing with zero real frequency. Depending on the parameters, these two modes may have comparable growth rates and the competition between the two has been a subject of interest \cite{Hellinger2000,Hellinger2001}. We here consider the FHI driven unstable by a temperature anisotropy of the EP component, which we call the EP-FHI. In other words, we consider a plasma consisting of the cold, isotropic ion and electron fluids with a low-density anisotropic EP population that provides the free energy. Linear analysis, as well as nonlinear simulation results, are presented below.

\begin{figure}[h!]
 \begin{center}
  \includegraphics[scale=0.5]{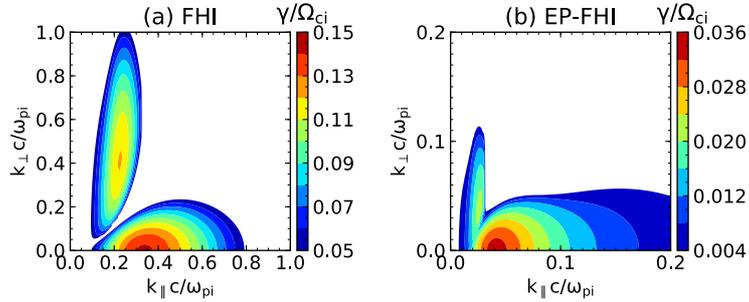}

  \caption{Growth rate for the firehose instability (FHI) driven by a parallel
  temperature anisotropy. The instability driven by (a) the thermal ion
  anisotropy (FHI), and (b) the energetic particle anisotropy (EP-FHI) are
  shown respectively. }

  \label{fig:parallel_growth}
 \end{center}
\end{figure}

Linear growth rates calculated with the full Vlasov-Maxwell equations as functions of the wavenumber for the FHI and EP-FHI are shown in Fig.~\ref{fig:parallel_growth}. An ion temperature anisotropy of $T_{\perp}/T_{\parallel} = 0.375$ and $\beta_{\parallel} = 4.0$ ($\beta_{\perp} = 1.5$) were used for the FHI shown in the left panel. The electrons were assumed to be isotropic and $\beta_e = 0.5$. \red Two peaks in the growth rate are clearly seen, one in parallel with a maximum growth rate $\gamma/\Omega_{ci} \simeq 0.14$ and another in oblique propagation with a growth rate of $\gamma/\Omega_{ci} \simeq 0.12$. \black We confirmed that the parallel propagating mode has a real frequency and the oblique mode is purely growing, consistent with the standard understanding. On the right panel, the growth rate for the EP-FHI is shown. The EP number density was assumed to be 1\% of the total density, whereas the EP temperature anisotropy and plasma beta were the same $T_{k,\perp}/T_{k,\parallel} = 0.375$ with $\beta_{k,\parallel} = 4.0$ and $\beta_{k,\perp} = 1.5$. Therefore, the result may reasonably be compared with the FHI. The plasma betas for the background ion and electron fluids were set to $\beta_i = \beta_e = 0.5$. We see that the result is qualitatively the same as the FHI. \red The growth rate of the parallel mode is $\gamma/\Omega_{ci} \simeq 0.034$ with a finite real frequency and the oblique mode is purely growing with a growth rate of $\gamma/\Omega_{ci} \simeq 0.021$. \black An important difference is that the wavenumbers for the unstable mode are shifted toward longer wavelengths. This may be understood by large Larmor radii of the EPs; a short wavelength mode with respect to the Larmor radii tends to be damped by the finite Larmor radius effect. The parallel and perpendicular thermal velocities for the EP population with the present parameters are $v_{k,\parallel}/v_A = \sqrt{\beta_{k,\parallel} (n_k/n_0)^{-1}) / 2} \simeq 14.1$, $v_{k,\perp}/v_A = \sqrt{\beta_{k,\perp} (n_k/n_0)^{-1} / 2} \simeq 8.67$. One may estimate the typical wavenumbers at which finite Larmor radius effect becomes important as $k_{\parallel} c/\omega_{pi} \approx (v_{k,\parallel}/v_A)^{-1} \simeq 0.07$ and $k_{\perp} c/\omega_{pi} \approx (v_{k,\perp}/v_A)^{-1} \simeq 0.11$. These estimates are roughly consistent with the high wavenumber cutoffs observed in the growth rates. The EP-FHI thus appears at the MHD scale, and accordingly, the frequency (for the parallel mode) and the growth rate are smaller than the FHI.

\red
We performed two nonlinear simulations with the standard hybrid and EP-MHD hybrid models, respectively. The standard hybrid simulation for FHI (Run-4A) employed the following parameters: $q_i/m_i = 0$, $q_k/m_k = -(m_e/m_i) q_e/m_e$, $\Omega_{ci} \Delta t = 0.02, \Delta h / c/\omega_{pi} = 0.5, N_x = N_y = 256, N_{ppc} = 256$. On the other hand, the parameters for the EP-MHD hybrid simulation for EP-FHI (Run-4B) were chosen as: $q_i/m_i = -(m_e/m_i) q_e/m_e$, $q_k/m_k = q_i/m_i$, $\Delta h / c/\omega_{pi} = 5.0$. The other parameters were the same as Run-4A. The spatial resolutions were chosen such that the most unstable modes are resolved by approximately the same number of grid points. We confirmed that the results remain essentially the same by increasing the spatial resolution (with correspondingly smaller box sizes).

\begin{figure}[h!]
 \begin{center}
  \includegraphics[scale=0.8]{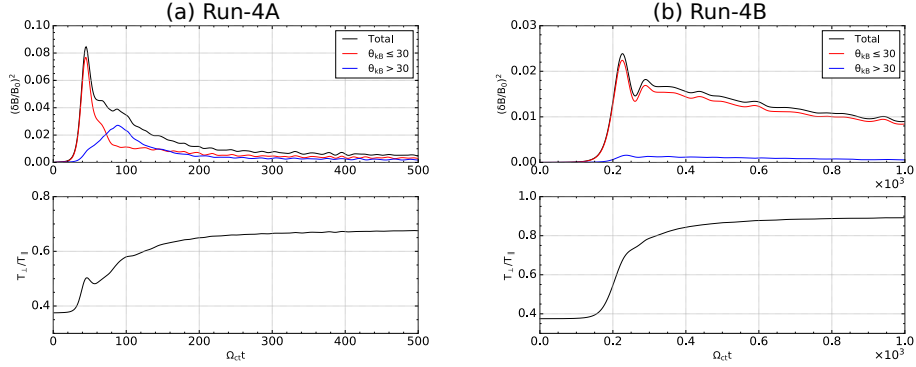}

  \caption{Comparison between (a) Run-4A and (b) Run-4B. The time evolution of
  magnetic field fluctuation amplitude (top) and temperature anisotropy
  (bottom) are shown. In the top panel, the black line shows the total wave
  amplitude. The red and blue lines show the amplitude integrated over
  wavenumber space in the ranges $\theta_{kB} \leq 30 \degree$ and
  $\theta_{kB} > 30 \degree$, respectively. Here, $\theta_{kB}$ is the angle
  between the wavenumber and the ambient magnetic field.}

  \label{fig:parallel_history}
 \end{center}
\end{figure}

Fig.~\ref{fig:parallel_history} compares results of Run-4A and Run-4B. In the top panel, the time evolution of the total wave energy $\delta B^2$ (black), and those integrated in wavenumber space over the regions $\theta_{kB} \leq 30 \degree$ (red) and $\theta_{kB} > 30 \degree$ (blue) are shown, respectively. Here $\theta_{kB}$ is the angle between the wavenumber and the ambient magnetic field direction, which was determined by taking the Fourier transform in space for each snapshot.

We see that the nonlinear evolution of the FHI (Run-4A) was quite similar to the results already presented in the literature \cite{Hellinger2001,Munoz2018}. Namely, the parallel FHI dominated in an early nonlinear phase whereas the power of the oblique FHI became comparable and eventually larger than the parallel in a late nonlinear phase. The reduction of anisotropy during the parallel FHI dominated phase ($\Omega_{ci} t \simeq 50$) was not strong enough and the oblique FHI continued to grow into larger amplitudes. The rate of pitch-angle scattering decreased after the oblique FHI saturated and its power started to decline. Eventually, $\sim 100 \Omega_{ci}^{-1}$ after the saturation, both of the parallel and oblique modes ended up with comparable amplitudes. We refer the reader to the earlier studies \cite{Hellinger2000,Hellinger2001} for more detailed discussion.
 
In contrast to the FHI, the results of the EP-FHI (Run-4B) as reproduced by the EP-MHD hybrid model was strikingly different. It is clear that the total wave energy was always dominated by nearly parallel modes. This may be partly because of different ratios of two growth rates (i.e., the parallel-to-oblique ratio of the growth rate is $\sim 1.2$ for the FHI and $\sim 1.6$ for the EP-FHI). However, the saturation of the oblique modes at the same time as the parallel, which was not the case in the FHI, indicates that the difference may not be explained by the linear property alone. Even without much of oblique-mode power, the temperature anisotropy reduced quite efficiently at around the saturation phase.

\begin{figure}[h!]
 \begin{center}
  \includegraphics[scale=0.5]{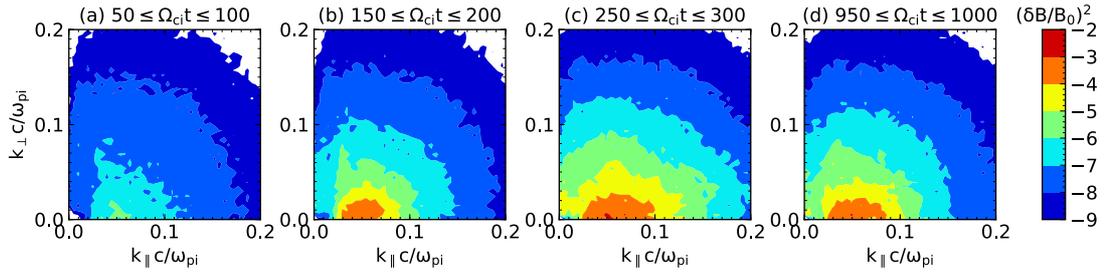}

  \caption{Wavenumber spectra of magnetic field fluctuations averaged during
  four time intervals: (a) $50 \leq \Omega_{ci} t \leq 100$, (b) $150 \leq
  \Omega_{ci} t \leq 200$, (c) $250 \leq \Omega_{ci} t \leq 300$, (d) $950
  \leq \Omega_{ci} t \leq 1000$ obtained by Run-4B. The base-10 logarithm of
  the power is shown as a function of the wavenumber. Note that only the first
  quadrant is shown due to the symmetry with respect to $k_{\parallel}$ and
  $k_{\perp}$ axes.}

  \label{fig:parallel_spectrum}
 \end{center}
\end{figure}

The wavenumber spectra of the magnetic field fluctuations averaged during four time intervals are shown in  Fig.~\ref{fig:parallel_spectrum}. These spectra were obtained by taking the Fourier transform and then averaged over the time intervals indicated in the figure: (a) $50 \leq \Omega_{ci} t \leq 100$, (b) $150 \leq \Omega_{ci} t \leq 200$, (c) $250 \leq \Omega_{ci} t \leq 300$, (d) $950 \leq \Omega_{ci} t \leq 1000$. The strongest peak in the early phase found in the parallel propagation is well consistent with the linear theory. There appeared a weak signature for the oblique mode ($k_{\parallel} c/\omega_{pi} \sim 0.03$, $k_{\parallel} c/\omega_{pi} \gtrsim 0.03$). Again, the weaker power of the oblique mode at the saturation of the parallel mode itself may be understood by the relatively small growth rate. However, the oblique modes no longer grew after the saturation of the parallel modes, as observed in the integrated power (Fig.~\ref{fig:parallel_history}).
\black

\begin{figure}[h!]
 \begin{center}
  \includegraphics[scale=0.5]{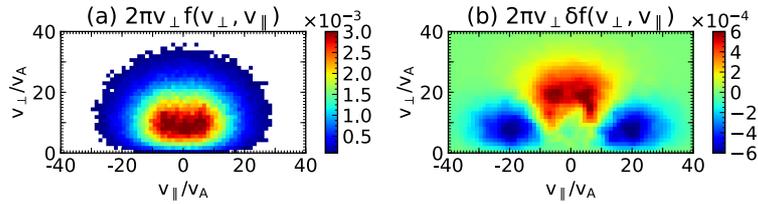}

  \caption{Snapshot of the velocity distribution function obtained by Run-4B
  integrated over the entire simulation box: (a) $2 \pi v_{\perp} f(v_{\perp},
  v_{\parallel})$ at the final snapshot $\Omega_{ci} t = 1000$, (b) the
  difference from the initial condition $2 \pi v_{\perp} \delta f(v_{\perp},
  v_{\parallel})$ with the same format as (a). The velocity distribution
  function is normalized such that the integral over velocity space becomes
  unity.}

  \label{fig:parallel_vdist}
 \end{center}
\end{figure}

To understand the saturation and the reduction of the anisotropy, we investigated the evolution of the velocity distribution function. Fig.~\ref{fig:parallel_vdist} shows the velocity distribution function $2 \pi v_{\perp} f(v_{\perp}, v_{\parallel})$ at the final stage $\Omega_{ci} t = 1000$ (the left panel) and the difference from the initial condition $2 \pi v_{\perp} \delta f(v_{\perp}, v_{\parallel})$ (the right panel), both integrated over the entire simulation box. The reduction of anisotropy was indeed associated with pitch-angle scatterings of small pitch angle particles ($\lesssim 45 \degree$) to larger pitch angles ($\gtrsim 45 \degree$). We confirmed that the efficient pitch-angle scatterings started to play the role in the saturation phase.

\red
One of the reasons that cause the discrepancy between the FHI and EP-FHI may be different dispersion characteristics and the resonance conditions. In the FHI, the parallel mode is on the whistler branch at relatively short wavelengths $k_{\parallel} c/\omega_{pi} \lesssim 1$. Therefore, the corresponding real frequencies are relatively high ($\omega/\Omega_{ci} \simeq 0.2$ for the wavenumber at the maximum growth rate). On the other hand, the oblique mode has zero real frequency. Consequently, the cyclotron resonance factors $\zeta_k = (\omega + \Omega_{ci})/\sqrt{2} k_{\parallel} v_{k,\parallel}$ as determined from the linear analysis are different between the two modes: $\zeta_k \simeq 1.9$ and $\zeta_k \simeq 1.5$ for the parallel and oblique modes, respectively. This somehow separates the dynamics between the parallel and oblique modes.

On the other hand, because of the finite Larmor radius effect associated with the EPs, the waves generated in the EP-FHI are long wavelength MHD waves. The wave frequency is thus much lower than the ion cyclotron frequency $\Omega_{ci}$ and the resonance factors for both modes are essentially the same $\zeta_k \simeq 1.3$. Therefore, if the initially dominant parallel modes scatter the particles in pitch angle via cyclotron resonance, the slope of the velocity distribution function that excites the instability will be smeared out. In other words, the free energy for the oblique mode is consumed by the parallel mode before it grows into larger amplitude. Also, since the resonance factor is smaller than the FHI, the major part of the distribution will be strongly affected by the resonant wave-particle interaction. Indeed, the reduction in anisotropy in the EP-FHI was more efficient than in the FHI.
\black

This provides an example where the presence of a dense cold plasma population changes the behavior of the instability driven by a pressure-carrying population. As we have seen here, the EPs tend to excite an instability at the MHD scale because of their relatively large Larmor radii.  Consequently, the self-consistent coupling between the MHD and the EP dynamics becomes important in general.

\section{Summary}
\label{sec:summary}

In this paper, we have proposed a new generalized hybrid model under the assumption of quasi-neutrality, which combines the standard hybrid model and the QNTF equations in a self-consistent manner. In other words, the background ion and electron fluids with arbitrary numbers of kinetic species comprise the system. The kinetic species whose dynamics is governed by the Vlasov equation may have arbitrary mass and energy densities, as well as charge-to-mass ratios. In the limit where the kinetic species dominates both the mass and energy densities, it reduces to the standard hybrid model with finite electron inertia effect. In the opposite limit where the presence of the kinetic species is negligible, the set of equations becomes identical to the QNTF equations. In an intermediate regime where the mass density is negligible whereas the energy density is substantial, the system (in the long wavelength limit) becomes the EP-MHD hybrid model. In this regime, the MHD and the EP dynamics are self-consistently coupled with each other.

A 3D numerical simulation code has been developed for the proposed equations. The effect of the kinetic species is included in the fluid equations written in the conservative form as the source terms. The kinetic species are updated with the standard PIC scheme. The electric field is determined by the generalized Ohm's law without any approximations which thus takes into account the contributions from all the species in the system. The magnetic field evolves according to Faraday's law. Since the code is based on the previously proposed HLL-UCT scheme for the QNTF equations, it automatically guarantees the divergence-free condition for the magnetic field. Although the scheme involves relatively large numerical dissipation, this makes the code quite robust in terms of numerical stability.

The simulation results presented in this paper have shown that the code correctly includes the kinetic effect associated with the kinetic species. We have also shown that by setting the charge-to-mass ratio of the ion fluid to zero ($q_i/m_i \rightarrow 0$), the same code may be used as a standard hybrid code with finite electron inertia effect. In the EP-MHD regime, the fluid assumption for the cold fluid reduces the computational costs and artificial noise. The required spatial resolution would be a fraction of the EP Larmor radius, which may be much larger than the ion inertial length or the Larmor radius of the thermal ions. Overall, the computational cost may be substantially reduced.

It is perhaps important to mention that the code provides a unified way to deal with vastly different parameter regions at the same time. For instance, a single simulation box may contain two distinct regions; one essentially a void of the kinetic species, another dominated by the kinetic species. The former automatically behaves as the MHD region, whereas the latter becomes the standard hybrid region. The interface region seamlessly connects the two regions without any artificial boundaries. This is a unique feature of the present model.

\blue
Currently, the code is parallelized using the MPI (Message Passing Interface) library with the standard three-dimensional domain decomposition. Since the load balancing capability has not been implemented yet, if there is significant inhomogeneity in the particle distribution (like the case mentioned above), the domain decomposition in the inhomogeneous direction should not be efficient. Although the computational time spent for the particle update is dominant whenever the number of particles is greater than, say, 100 particles/cell, the coexistence of the fluid dominant and particle dominant regions in the same simulation domain makes computational load balancing rather difficult. This clearly needs further investigation.
\black

The critical disadvantage of the present model is that one has to resolve the fast gyromotion of the particles. For large scale problems with time scales much slower than the ion gyroperiod, this imposes a severe restriction on the time step. In principle, the problem may be circumvented by adopting the drift-kinetic or gyrokinetic approximations for the EP population. It is relatively easy if the kinetic species is a minor component such that the ideal MHD Ohm's law may be employed \cite{Todo1995,Belova1997}. However, for general situations, one has to be careful in including the contributions to the moment quantities in the generalized Ohm's law. They must be modified to be consistent with the approximation adopted. In any case, these approximations may be useful for practical applications where the time step restriction becomes the critical issue.

\section*{Acknowledgments}
This work was supported by JSPS KAKENHI Grant Numbers 16H01170, 17H02966, 17H06140, and also by the joint research program of the Institute for Space-Earth Environmental Research (ISEE), Nagoya University. Numerical computations were carried out on the Cray XC30 at Center for Computational Astrophysics, National Astronomical Observatory of Japan, and the CX400 supercomputer system at the Information Technology Center, Nagoya University.


%% trick for figure numbering
\let\thefigureSAVED\thefigure
\appendix
\let\thefigure\thefigureSAVED

\section{Derivation of Generalized Ohm's Law}
\label{appendix:ohmslaw}

Derivation of the generalized Ohm's law in the form of Eq.~(\ref{eq:ohm}) was given in \cite{Amano2014,Amano2015} for the standard hybrid and QNTF models. Here we reiterate essentially the same procedure explicitly for the present model for the sake of completeness.

\blue
We start with Maxwell's equations in the low-frequency approximation Eqs.~(\ref{eq:faraday}) and (\ref{eq:ampere}).  Taking temporal derivative of Ampere's law Eq.~(\ref{eq:ampere}) and making use of Faraday's law Eq.~(\ref{eq:faraday}) yield the following equation: \black
\begin{align}
 -\frac{c^2}{4 \pi} \nabla \times \nabla \times \bm{E} =
 \frac{\partial}{\partial t} \bm{J}.
\end{align}
The current density in the present system is given by the sum over all species Eq.(\ref{eq:current}). Therefore, the right-hand side of the above equation may be written as follows:
\begin{align}
 \frac{\partial}{\partial t} \bm{J}
 &=
 \frac{\partial}{\partial t}
 \left[
 \sum_{s=i,e} \frac{q_s}{m_s} \rho_s \bm{v}_s +
 \sum_{s=kinetic} q_s \int \bm{v} f_s d \bm{v}
 \right]
 \nonumber \\
 &=
 - \nabla \cdot
 \left[
 \sum_{s=i,e} \frac{q_s}{m_s}
 \left( \rho_s \bm{v}_s \bm{v}_s + p_s \bm{I} \right) +
 \sum_{s=kinetic} q_s \int \bm{v} \bm{v} f_s d \bm{v}
 \right]
 \nonumber \\
 &+
 \left(
 \sum_{s=i,e} \frac{q_s^2}{m_s^2} \rho_s +
 \sum_{s=kinetic} \frac{q_s^2}{m_s} \int f_s d \bm{v}
 \right) \bm{E}
 +
 \left(
 \sum_{s=i,e} \frac{q_s^2}{m_s^2} \rho_s \bm{v}_s +
 \sum_{s=kinetic} \frac{q_s^2}{m_s} \int \bm{v} f_s d \bm{v}
 \right) \times \frac{\bm{B}}{c}
 \nonumber \\
 &=
 \frac{1}{4 \pi}
 \left[
 \Lambda \bm{E} + \frac{\bm{\Gamma}}{c} \times \bm{B} -
 \nabla \cdot \bm{\Pi}
 \right]
\end{align}
where we have used Eqs.~(\ref{eq:basic_momentum}) and (\ref{eq:vlasov}) to replace the temporal derivatives. Note also the fact that the Lorentz force is divergence free with respect to the derivative by velocity. It is easy to confirm that $\Lambda$, $\bm{\Gamma}$, and $\bm{\Pi}$ correspond to the moment quantities as defined in Eqs.~(\ref{eq:lambda})-(\ref{eq:pi}). The generalized Ohm's law therefore can be written as
\begin{align}
 \left( \Lambda + c^2 \nabla \times \nabla \right) \bm{E} =
 - \frac{\bm{\Gamma}}{c} \times \bm{B}
 + \nabla \cdot \bm{\Pi},
\end{align}
which is equivalent to Eq.~(\ref{eq:ohm}) except for the resistive term. Since the above derivation does not involve any approximation, this form of the generalized Ohm's law is exact \blue as long as the initial assumptions (i.e., low-frequency approximation for Ampere's law and fluid approximation on the electrons and background ions) are correct. \black We have previously shown that the exact form of the Ohm's law correctly includes the finite electron inertia effect \cite{Amano2014,Amano2015}.

\section{Reduction to Standard Hybrid}
\label{appendix:standard_hybrid}

One has to obtain primitive variables $\rho_s$, $\bm{v}_s$, and $p_s$ for
the ion and electron fluids from the conservative variables
\begin{align}
 D &= \rho_i + \rho_e,
 \nonumber \\
 \bm{M} &= \rho_i \bm{v}_i + \rho_e \bm{v}_e
 \nonumber \\
 K &= \varepsilon_i + \varepsilon_e +
 \frac{\bm{B}^2}{8\pi}
\end{align}
every time step for calculating the numerical fluxes as well as for the
generalized Ohm's law.

One may determine the electron fluid density and velocity from
\begin{align}
 \rho_e &=
 \frac
 {\varrho_k + \frac{q_i}{m_i} D - \frac{1}{4 \pi} \nabla \cdot \bm{E}}
 {\frac{q_i}{m_i} - \frac{q_e}{m_e}}
 \label{eq:primitive_rhoe} \\
 \bm{v}_e &=
 \frac
 {\bm{J}_k + \frac{q_i}{m_i} \bm{M} - \frac{c}{4 \pi} \nabla \times \bm{B}}
 {\rho_e (\frac{q_i}{m_i} - \frac{q_e}{m_e})}.
 \label{eq:primitive_ve}
\end{align}
Because of the quasi-neutral assumption, $\nabla \cdot \bm{E} \approx 0$
is used in practice. The primitive variables for the ion fluid are then
readily obtained as
\begin{align}
 \rho_i &= D - \rho_e,
 \\
 \bm{v}_i &= \frac{\bm{M} - \rho_e \bm{v}_e}{\rho_i}
 \\
 p_i &= \left(\gamma - 1\right)
 \left(
 K
 - \frac{1}{2} \rho_i \bm{v}_i^2 - \frac{1}{2} \rho_e \bm{v}_e^2
 - \frac{\bm{B}^2}{8\pi}
 \right) - p_e,
\end{align}
where the electron pressure $p_e$ can be obtained using an equation of state,
for instance.

Now let us show that these equations automatically reduce to the standard
hybrid model by taking the limit $q_i/m_i \rightarrow 0$. The electron density
and velocity in this case are:
\begin{align}
 \rho_e
 &\rightarrow
 \frac{\varrho_k - \frac{1}{4 \pi} \nabla \cdot \bm{E}}{-\frac{q_e}{m_e}}
 = \frac{m_e}{e} \sum_{s=kinetic} \frac{q_s}{m_s} I_s^0,
 \\
 \bm{v}_e
 &\rightarrow
 \frac{\bm{J}_k - \frac{c}{4 \pi} \nabla \times \bm{B}}
 {- \rho_e \frac{q_e}{m_e}}
 =
 \frac{1}{n_e e}
 \left(
 \sum_{s=kinetic} \frac{q_s}{m_s} \bm{I}_s^1
 - \frac{c}{4\pi} \nabla \times \bm{B}
 \right),
\end{align}
where $e$ is the elementary charge. Note that $I_s^0$ and
$\bm{I}_s^1$ are defined by Eqs.~(\ref{eq:mom0}) and (\ref{eq:mom1}),
respectively. These are indeed identical to the equations used in a standard
hybrid code. The ion fluid variables have no effect on the generalized Ohm's
law because all the moment quantities are proportional to $q_i/m_i$ or
$(q_i/m_i)^2$. Therefore, exactly the same code can be used as a standard
hybrid code by setting $q_i/m_i = 0$.

%% References


\bibliographystyle{elsarticle-num}
\bibliography{reference}
%\input{ms.bbl}



%\begin{thebibliography}{00}



% \end{thebibliography}
\end{document}